# Designing topological interface states in phononic crystals based on the full phase diagrams

Yan Meng,[1,2,3,a)] Xiaoxiao Wu,[2,a)] Ruo-Yang Zhang,[2] Xin Li, [1,3] Peng Hu,[3] Lixin Ge,[3] Yingzhou Huang,[1,3] Hong Xiang,[3] Dezhuan Han,[3] Shuxia Wang,[1,3,b)] Weijia Wen[1,2,b)]

[1] *Soft Matter and Interdisciplinary Research Center, College of Physics, Chongqing University, Chongqing, 400044, P. R. China*

[2] *Department of Physics, The Hong Kong University of Science and Technology, Clear Water Bay, Kowloon, Hong Kong, China*

[3] *Department of Applied Physics, Chongqing University, Chongqing 401331, China*

[a)] These authors contribute equally.

[b)] Correspondence and requests for materials should be addressed to S.X.W. (email: wangshuxia@cqu.edu.cn) or to W.J.W. (email: phwen@ust.hk).

## Abstract

The topological invariants of a periodic system can be used to define the topological phase of each band and determine the existence of topological interface states within a certain bandgap. Here, we propose a scheme based on the full phase diagrams, and design the topological interface states within any specified bandgaps. As an example, here we propose a kind of one-dimensional phononic crystals. By connecting two semi-infinite structures with different topological phases, the interface states within any specific bandgap or their combinations can be achieved in a rational manner. The existence of interface states in a single bandgap, in all odd bandgaps, in all even bandgaps, or in all bandgaps, are verified in simulations and experiments. The scheme of full phase diagrams we introduce here can be extended to other kinds of periodic systems, such as photonic crystals and designer plasmonic crystals.

## Introduction

The topological physics is growing rapidly in condensed matter physics, from quantum Hall effect [1] to topological insulators [2,3] and Weyl semimetals [4]. Topological insulators possess topologically non-trivial bandgaps, and support one-way surface states. While Weyl semimetals possess Weyl nodes, which give rise to monopoles and anti-monopoles in the momentum space, and support topological Fermi-arc surface states. These topological surface states appear at the interface of two insulators with different topological phases, and therefore are robust against some local defects and immune from the back-scattering. In light of these advantages, recent discoveries of topological interface states have been extended to various physical branches, including photonics [5-14] and phononics [15-20].

For one-dimensional (1D) periodic systems, the simplest topologically nontrivial phase exists in polyacetylene [21]. It is found that the band topology of this kind of systems can be characterized by Zak phases, which are quantized topological invariants as long as the unit cell possesses inversion symmetry [22]. In recent years, the topological description has been introduced in various classical counterpart such as 1D photonic crystals and phononic crystals (PCs) [23, 24], and been successfully applied to predict the existence of interface states from the Zak phases of bulk bands.

In the system of a 1D PC, for example, a cylindrical waveguide with periodically alternating structures, the macroscopic controllability enables it to be a capable platform to realize the advanced concepts such as band inversion and topological phase transition. Recently, interesting topological phenomena, such as topological interface and edge states [24-26], valley-Hall effect [27], have been observed in acoustic systems.

In this work, we focus on the topologically induced interface states within arbitrary bandgap (or bandgaps) by judiciously designing the Zak phases of each constituent PC. The existence of interface states can be equivalently predicted by the Zak phases, surface impedances, and transmission spectra. As an example, all possible existence of the interface states in the lowest four bandgaps is observed numerically and experimentally, including the interface states in any single bandgap, all odd bandgaps, all even bandgaps, and all bandgaps. The transmission spectra and spatial distributions of the pressure field for the interface states are measured and exhibit excellent agreement with the simulated results.

**Results**

**Topological properties of 1D PCs.** The 1D periodic system under study is shown in Fig. 1a. The PC is composed of two kinds of sound hard tubes of different sizes, and filled with air. A unit cell of this PC is shown in Fig. 1b, and the corresponding geometric parameters are labeled. The unit cell is a bipartite combination of two alternate tubes (tube-A and tube-B) with different radii ($r_A$, $r_B$) and lengths ($d_A$, $d_B/2$), and possesses inversion symmetry where the inversion center can be selected as the middle of either tube-A or tube-B.

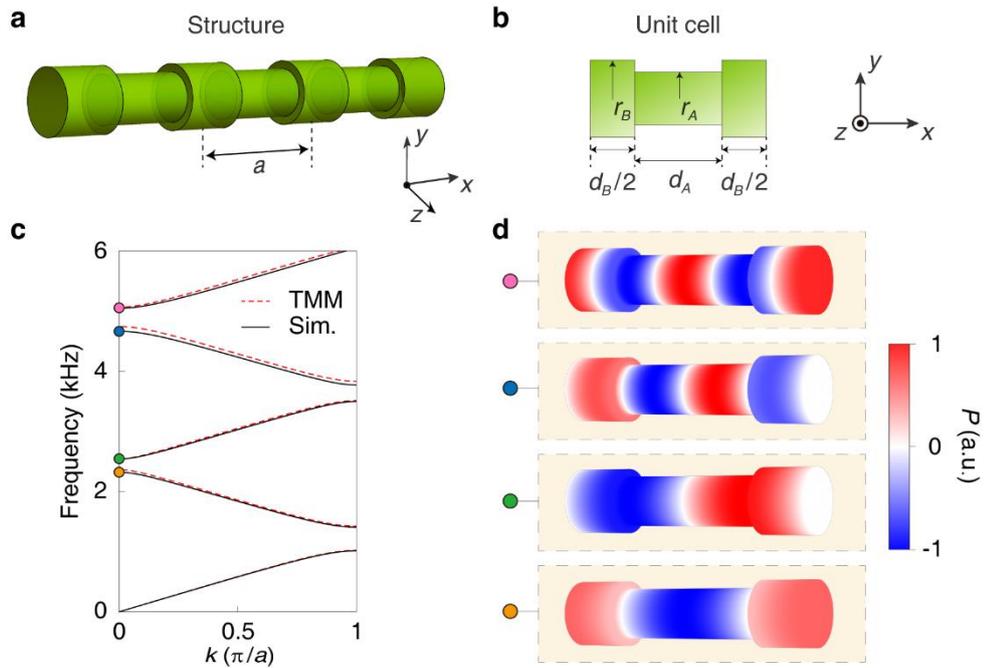

**Figure 1| One-dimensional phononic crystal system. a**, Schematic illustration of the one-dimensional (1D) phononic crystal (PC) consisting of alternating tubes with different sizes. A unit cell of the 1D PC is marked by dashed lines with a lattice constant *a*. **b**, Side-view of the unit cell. The unit cell is centered at the middle of tube-A with inversion symmetry. The geometric parameters are labelled correspondingly. **c**, Band structures calculated by the transfer matrix method and numerical simulation are plotted by red dashed and black solid lines, respectively. The dimensions for this

structure (denoted by $S_1$) are $r_A$=1.3 cm, $r_B$=1.7 cm, $d_A$=8 cm, $d_B$=6 cm. The frequencies of the lowest four band-edge modes at $k$=0 are indicated by orange (2.321 kHz), green (2.541 kHz), blue (4.664 kHz) and pink (5.046 kHz) dots, respectively. **d**, Simulated eigen-fields of pressure in the unit cell of $S_1$, corresponding to the four labeled band-edge modes in **c**. These four eigenmodes can be classified into the even modes (first and last) or odd modes (second and third) according to the field distributions with respect to the inversion center.

The band structure of the 1D PC with sound hard boundaries can be obtained by the transfer matrix method (TMM) [28, 29],

$$\cos(ka) = \cos\left(\frac{\omega a}{v_a}\right) - \frac{1}{2}\left(\frac{S_A}{S_B} + \frac{S_B}{S_A} - 2\right)\sin\left(\frac{\omega d_A}{v_a}\right)\sin\left(\frac{\omega d_B}{v_a}\right), \quad (1)$$

where $k$ is the Bloch wave vector, $\omega$ is the angular frequency, $v_a$ is the sound speed in air (343 m/s), $a$ is the lattice constant ($a=d_A+d_B$), and $S_{A(B)} = \pi r_{A(B)}^2$ is the cross-sectional area for tube-A(B). As shown in Fig. 1c, the acoustic band structure obtained by TMM (red dashed) agrees well with the full-wave simulated one (black solid). The geometric parameters of the unit cell are labeled in Fig. 1b. The dimensions here are chosen as $r_A$=1.3 cm, $r_B$=1.7 cm, $d_A$=8 cm, $d_B$=6 cm. The PC constructed with these geometric parameters is denoted by $S_1$.

The pressure field distribution of the eigenmode in the PC can be written as,

$$P_{n,k}(x,\mathbf{r}) = u_{n,k}(x,\mathbf{r})e^{ikx}, \quad (2)$$

where $u_{n,k}(x,\mathbf{r})$ is the periodic-in-cell part of the Bloch eigenfunction of the state on the $n$-th band with the wave vector $k$, $x$ is the coordinate along the axial direction, $\mathbf{r}$ is the position vector in the transverse ($y$-$z$) plane. When the wavelength $\lambda$ satisfies $\lambda/4$>Max$\{r_A, r_B\}$ (See Supplementary Note 1), no transverse mode can be supported in the PC, and $P_{n,k}(x,\mathbf{r})$ will be nearly uniform in the transverse cross section. Therefore, the PC can be considered as a 1D system here. As examples, four band-edge modes at the Γ point ($k$=0) are indicated by differently colored dots in Fig. 1c, and their numerically simulated eigen-fields are shown in Fig. 1d correspondingly. As illustrated in Fig. 1d, the pressure field for each eigenmode possesses either even or odd parity with respect to the center of the unit cell, while is approximately uniform in the cross-sectional plane.

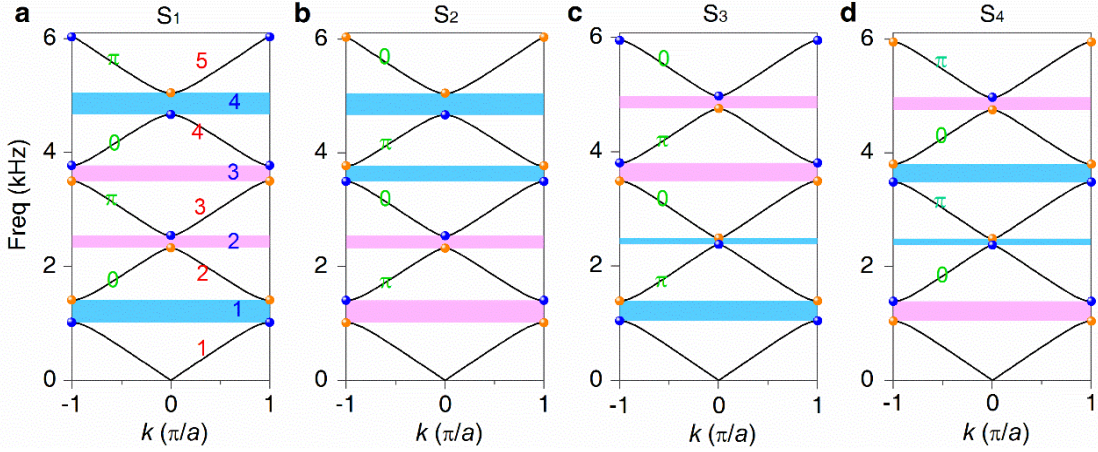

**Figure 2| Topological properties for the pass bands and bandgaps. a-d**, The band structures for $S_1$ ($\delta r=-0.2$ cm, $\delta d=1$ cm), $S_2$ ($\delta r=0.2$ cm, $\delta d=-1$ cm), $S_3$ ($\delta r=-0.2$ cm, $\delta d=-1$ cm), and $S_4$ ($\delta r=0.2$ cm, $\delta d=1$ cm), respectively. Here, $S_1$ and $S_2$ ($S_3$ and $S_4$) share the same PC structure but possess different choice of the unit cell centered respectively at tube-A and tube-B. The signs of the imaginary part of surface impedance $\zeta$ ($Z/Z_0=i\zeta$) for the gaps are indicated by the magenta strips for $\zeta>0$, and cyan strips for $\zeta<0$. The Zak phase for each band is shown in green, and the ordinals of bands and gaps are listed respectively by the red and blue labels. The blue and orange dots represent the band-edge states with odd and even parities, respectively.

The topological property of bulk bands for the 1D PC can be represented by their Zak phases, which are defined as [24]:

$$\theta_n^{Zak} = \int_{-\pi/p}^{\pi/p} dk \left[ i \int_{\text{unit cell}} \frac{1}{2\rho v_a^2} d\mathbf{r} dx u_{n,k}^*(x,\mathbf{r}) \partial_k u_{n,k}(x,\mathbf{r}) \right], \quad (3)$$

where $\theta_n^{Zak}$ is the Zak phase of $n^{\text{th}}$ bulk band, and $\rho$ is the density of air. In general, $\theta_n^{Zak}$ can be any value if the choice of the unit cell is arbitrary. However, when the unit cell is chosen to be inversion symmetric with the inversion center being the middle of tube-A or tube-B, it can be proved that $\theta_n^{Zak}$ should be quantized as either 0 or $\pi$. The Zak phase can take different values (0 or $\pi$) for different choices of the unit cell [22, 23]. The quantized $\theta_n^{Zak}$ characterizes the topology of the corresponding band.

When two semi-infinite PCs are connected at an interface, the condition for the existence of an interface state in the $n^{th}$ bandgap is that the impedances on both sides of the interface satisfy the condition: $Z_L+Z_R=0$ [23]. Here, $Z_L$ and $Z_R$ are respectively the impedances of the left-hand and right-hand PCs, and can be expressed by the reflection coefficient from the left (right) side of the interface as $Z_{L(R)}=Z_0[1+r_{L(R)}]/[1-r_{L(R)}]$, where $Z_0$ is the impedance of the free space. As is known, for the semi-infinite phononic crystal, the surface impedance $Z$ is purely imaginary in the bandgap, i.e. $Z/Z_0=i\zeta$. According to the bulk-interface correspondence for the 1D PC, the sign of $\zeta^{(n)}$ within the $n^{th}$ gap can be related to the Zak phases of the bulk bands below this gap by the following relation, as long as there is no band crossing below this gap [23]:

$$\text{sgn}[\zeta^{(n)}] = (-1)^{n+1}\exp\left(i\sum\nolimits_{m=1}^{n}\theta_m^{Zak}\right). \tag{4}$$

Since the condition $Z_L+Z_R=0$ can be satisfied if $\zeta$ takes opposite sign for the left-hand and right-hand PCs, we can directly predict the existence of interface state inside a common bandgap of the two connected PCs from the information of their Zak phases.

The topological properties of acoustic band structure can be easily manipulated by tuning the geometric parameters of PC. To construct a parameter space, two variables $\delta r\equiv(r_A-r_B)/2$ and $\delta d\equiv(d_A-d_B)/2$ are adopted, while $r_m=(r_A+r_B)/2=1.5$ cm and $d_m=(d_A+d_B)/2=7$ cm are fixed. In this parameter space, we choose other three structures as $S_2$ ($\delta r=0.2$ cm, $\delta d=-1$ cm), $S_3$ ($\delta r=-0.2$ cm, $\delta d=-1$ cm), and $S_4$ ($\delta r=0.2$ cm, $\delta d=1$ cm) to compare with $S_1$. These four structures share the common frequency intervals for all of their lowest four bandgaps.

The band structures for $S_1$, $S_2$, $S_3$ and $S_4$ are shown in Figs. 2a-2d, respectively. The corresponding Zak phase for each individual band is numerically calculated (see method) and denoted by the green letter. Meanwhile, the sign of $\zeta^{(n)}$ is indicated by the magenta (positive) or cyan (negative) strips, respectively. The sign of $\zeta^{(n)}$ can be identified not only by Eq. (4) but also by the parity of the band-edge states below and above the corresponding bandgap [23]. The band-edge states are marked by the dots at the Brillouin zone boundary ($k=\pm\pi/a$) and center ($k=0$), and their parities are represented by their colors with blue for odd and orange for even respectively to the odd and even modes. If the eigenmode at the lower edge of the bandgap has odd parity while the one at upper edge is even, then sgn($\zeta$)<0; otherwise, if the eigenmode at lower edge is even and the one at the upper edge is odd, then sgn($\zeta$)>0 [23].

Topologically induced interface states in $n^{th}$ gap can be achieved by different combinations of structures $S_i$ and $S_j$ ($i, j$=1, 2, 3, 4, and $i \neq j$) as long as $\zeta^{(n)}$ of the left-hand and right-hand PCs take opposite sign, i.e. the colors in the $n^{th}$ bandgap for $S_i$ and $S_j$ are different. As shown in Fig. 2, for the configuration of $S_1$+$S_2$ (or $S_3$+$S_4$), there should exist interface states in all the odd bandgaps. Whereas for the configuration of $S_1$+$S_3$ (or $S_2$+$S_4$), there are interface states in the even bandgaps. For $S_1$+$S_4$ (or $S_2$+$S_3$), the two PCs fall into the opposite phases in all bandgaps, consequently, interface states appear in all the bandgaps. Note that when the system is impinged by a plane wave, the inference for these 3 cases is not restricted to the first four bandgaps, but is applicable for all higher-ordinal bandgaps as long as the weak dispersion limit is satisfied [30].

**Achieving interface states from the full phase diagram.** Analogous to the band inversion in electronic systems [31, 32], when the bandgap closes and reopens implies that the topological phase transition occurs in the system of PC. The loci of the band crossing point will separate the different phases of a certain bandgap, which can be characterized by the sign of $\zeta^{(n)}$ within this bandgap. With the condition of band crossing [23, 30], band inversion can be achieved when the geometric parameters in Eq. (1) satisfy: (i) $r_A$=$r_B$, then the structure degenerates to the trivial case, i.e. a non-structured tube with a constant cross section; or (ii) $d_A/d_B$=$(d_m+\delta d)/(d_m-\delta d)$=$n_1/n_2$, where $n_1$ and $n_2$ are integers, as such the band crossing occurs at the $(n_1+n_2)^{th}$ bandgap [23]. According to these two criterions, all the phase transition lines in the phase diagram of a bandgap can be obtained.

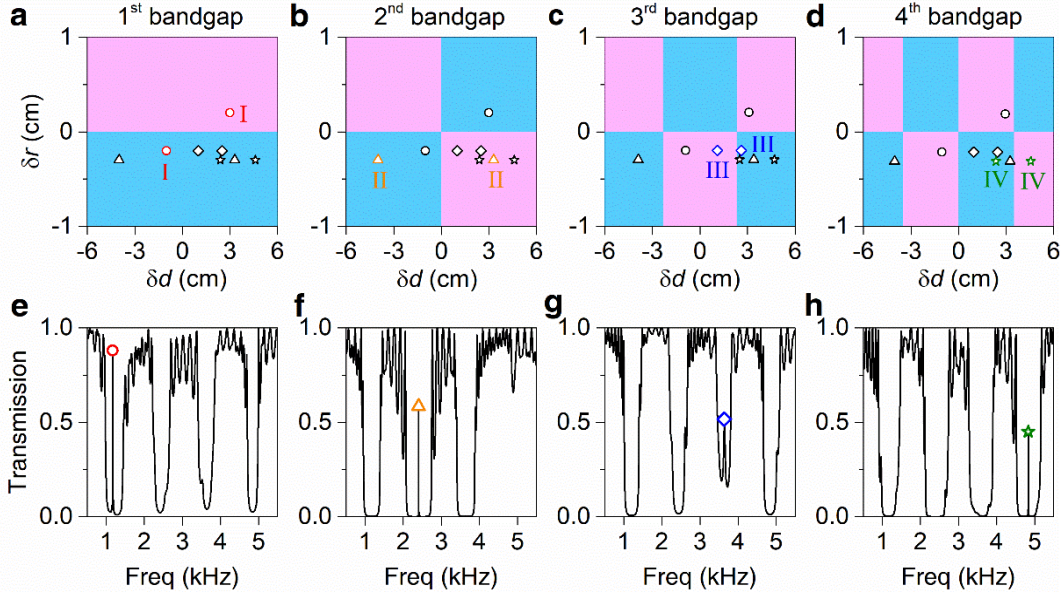

**Figure 3| Constructing interface states in any single bandgap from the phase diagrams in the ($\delta r$, $\delta d$) space. a-d,** The phase diagrams of the lowest 4 bandgaps. The phases of sgn($\zeta$)>0 and sgn($\zeta$)<0 are colored in magenta and cyan, respectively. The four pairs of geometric parameters, marked by circles (I), triangles (II), rhombus (III), and stars (IV), respectively, are chosen to construct four configurations of connected PCs. For each pair, the two points in the parameter space belong to different phases in only one of the four gaps and are highlighted by different colors. The corresponding simulated transmission spectra for the four configurations of connected PCs are plotted in **e-h**. For each configuration, there is only one peak corresponding to an interface state in the $n^{th}$ (n=1,2,3,4) bandgap, which are marked by a red circle, orange triangle, blue rhombus, and green star, respectively.

For the geometric parameters $r_m$ and $d_m$ chosen here, the condition $\lambda/4 > \text{Max}\{r_A, r_B\}$ is satisfied in the frequency interval for the lowest four bandgaps. As such, only these four bandgaps are considered in the following. The full phase diagrams of the 4 bandgaps based on the TMM approximation are shown in the upper panel of Fig. 3. The phases of sgn($\zeta$)>0 and sgn($\zeta$)<0 are colored in magenta and cyan, respectively. Except for the line of $\delta r=0$, the first bandgap has no other phase transition lines, while the second bandgap is also closed at $\delta d=0$, the third bandgap is closed at $\delta d=\pm d_m/3$, and the fourth bandgap is closed at $\delta d=0$, $\delta d=\pm d_m/2$.

Considering two connected semi-infinite PCs denoted by $S_L$ and $S_R$, the sufficient condition for the existence of an interface state within the $n^{th}$ bandgap is that $S_L$ and $S_R$ drop into different phases in the corresponding phase diagram. Therefore, the phase diagram can give us the complete information to construct the geometric structures of two connected PCs and form an interface in the $n^{th}$ bandgap. For example, we can choose the geometric parameters intentionally to construct the connected PCs such that the interface state only exists in a single bandgap. As shown in the upper panel of Fig. 3, the geometric parameters for the connected PCs ($S_L$ and $S_R$) are marked by circles, triangles, rhombus, and stars, respectively. In the phase diagram of the first bandgap, the two circles (marked in red) are located in the regions of different phases; for the other three bandgaps, however, the circles belong to the same phases. Accordingly, the topological interface state exists only in the first bandgap for the connected PCs: $S_{L,I}+S_{R,I}$. The corresponding simulated transmission spectrum shown in Fig. 3(e) verifies this prediction. Similarly, for other three pairs of PCs marked by triangles, rhombus, and stars, the phases of left-hand and right-hand PCs are only different in the 2$^{nd}$, 3$^{rd}$ and 4$^{th}$ bandgaps, respectively, thus there should exist only one interface state within the corresponding bandgap for each configuration. The corresponding transmission spectra shown in Figs. 3(f), 3(g) and 3(h) confirm the existence of such interface states.

| Table1\| signs of $\zeta$ for four bandgaps in the parameter space | | | | | | |
|---|---|---|---|---|---|---|
| $\delta r$ \ $\delta d$ | a | b | c | d | e | f |
| A | 1,2,3,4 | 1, 2, 3, −4 | 1,2, −3, −4 | 1, −2, −3,4 | 1, −2, 3, 4, | 1, −2,3, −4 |
| B | −1, −2, −3, −4 | −1 −2 −3 4 | −1, −2, 3,4 | −1,2, 3, −4 | −1, 2, −3, −4, | −1,2, −3,4 |

> **a-f** represent the range in the $\delta d$-axis, and **a**: $-6<\delta d<-3.5$, **b**: $-3.5<\delta d<-7/3$, **c**: $-7/3<\delta d<0$, **d**: $0<\delta d<7/3$, **e**: $7/3<\delta d<3.5$, **f**: $3.5<\delta d<6$.
> **A-B** represent the range in the $\delta r$-axis, and **A**: $0<\delta r<1$, **B**: $-1<\delta r<0$.
> The unit here is cm. The positive (negative) number denotes the bandgap with positive (negative) $\zeta$.

With the full phase diagrams in the ($\delta r$, $\delta d$) space, we can construct topologically induced interface states in arbitrary bandgap(s). As shown in Table 1, the parameter space is divided into twelve blocks labeled by the vertical (**A**, **B**) and horizontal (**a-f**) indices. The signs of $\zeta$ within the $n^{th}$ bandgap are listed in each block. The positive (negative) number denotes the bandgap with positive (negative) $\zeta$. When the geometric parameters of two connected PCs are located in the same block, the signs of $\zeta$ for the two PCs are identical in all of the four gaps; accordingly, there are no topological interface state since no topological phase transition take places. However, when the geometric parameters of two PCs are located in different blocks, the numbers in each block can be regarded as a set, and the symmetric difference between these two sets can indicate within which gap(s) the sgn($\zeta$) of the two PCs are opposite. Thus with the full phase diagram, we can predict whether there is an interface state within each bandgap for a certain combination of two PCs. For instance, consider the combination of two PCs with geometric parameters located in **Aa** and **Bc** blocks, and the symmetric difference can be denoted as $\mathbf{Aa}\Delta\mathbf{Bc} = \left\{ n \mid \text{sgn}\left(\zeta_{\mathbf{Aa}}^{(n)}\right) \neq \text{sgn}\left(\zeta_{\mathbf{Bc}}^{(n)}\right) \right\}$, and the result is {1, 2}, consequently, the interface states can only appear in the 1$^{st}$ and 2$^{nd}$ bandgaps. In terms of this simple method, we list all possible combinations of two connected PCs and the corresponding interface states for the lowest 4 gaps (see Supplementary Note 2).

The numerical results of band inversion in the ($\delta r$, $\delta d$) space are shown in Fig. 4. The upper and lower bounds of each bandgap are illustrated in red or blue, where the colors indicate the parity of the band-edge modes (red for even, blue for odd). The intersection of the two surfaces corresponds to the phase transition lines in the corresponding bandgap which are shown by magenta and cyan in the projected plane.

Qualitative agreement of the phase diagrams for each bandgap can be observed from the numerical and TMM results (see details in Supplementary Note 3). Therefore, the phase diagrams calculated from TMM are satisfactorily accurate and can be used to construct topologically induced interface states in our system.

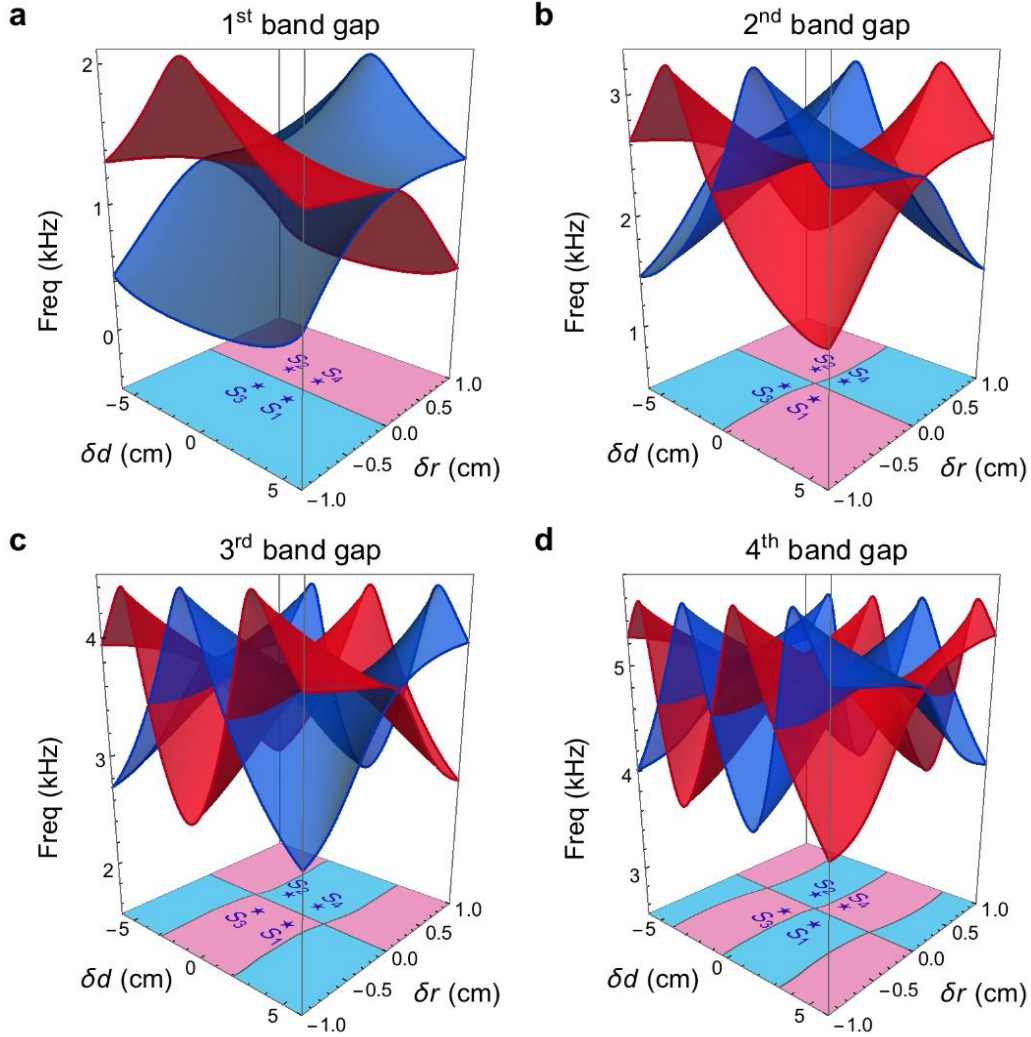

**Figure 4| Topological phase diagrams in the 1D PC system. a-d**, The frequency intervals of the 1st, 2nd, 3rd, and 4th bandgaps change in the parameter space, where the red and blue surfaces correspond to the band-edge frequencies with even and odd parities, respectively. The intersection of the two surfaces corresponds to the transition lines of the two topological phases, namely sgn($\zeta$)>0 and sgn($\zeta$)<0, which are shown by magenta and cyan in the projected plane. The geometric parameters for the configurations $S_1$-$S_4$ are marked by the blue stars in the projected plane.

**Experimental observation of the designed interface states.** To confirm the theoretical prediction of topological interface states from the phase diagrams, numerical simulations and experimental measurements have been performed. The reflection coefficient $r_{Si}$ for the structure $S_i$ ($i$=1 to 4) is simulated with 80 units for each structure. Note that the above mentioned condition for the presence of interface states: $Z_{S1}+Z_{Si}=0$ is equivalent to $r_{S1}r_{Si}=1$. The products $r_{S1}r_{S2}$, $r_{S1}r_{S3}$ and $r_{S1}r_{S4}$ are shown in Figs. 5(b)-4(d), and the real (imaginary) part of $r_{S1}r_{Si}$ is plotted in black (gray). The condition $r_{S1}r_{Si}=1$ can be satisfied in the 1$^{st}$ and 3$^{rd}$ bandgaps for the $S_1+S_2$ configuration, in the 2$^{nd}$ and 4$^{th}$ gaps for the $S_1+S_3$ configuration, and in all of the 4 gaps for the $S_1+S_4$ configuration. These results verify the prediction based on the phase diagrams. The simulated transmission spectra for these three configurations with 6 unit cells on both sides of the interface are exhibited in Figs. 5(e)-(g). The transmission peaks marked by colored dots in each bandgap, unequivocally confirm the existence of the interface states at the theoretically predicted frequencies. Further simulations reveal that the influence of the finite size effect of the two PCs on the transmission spectra is less than 5% as long as the number of unit cells is larger than 4 (further details are given in Supplementary Note 4).

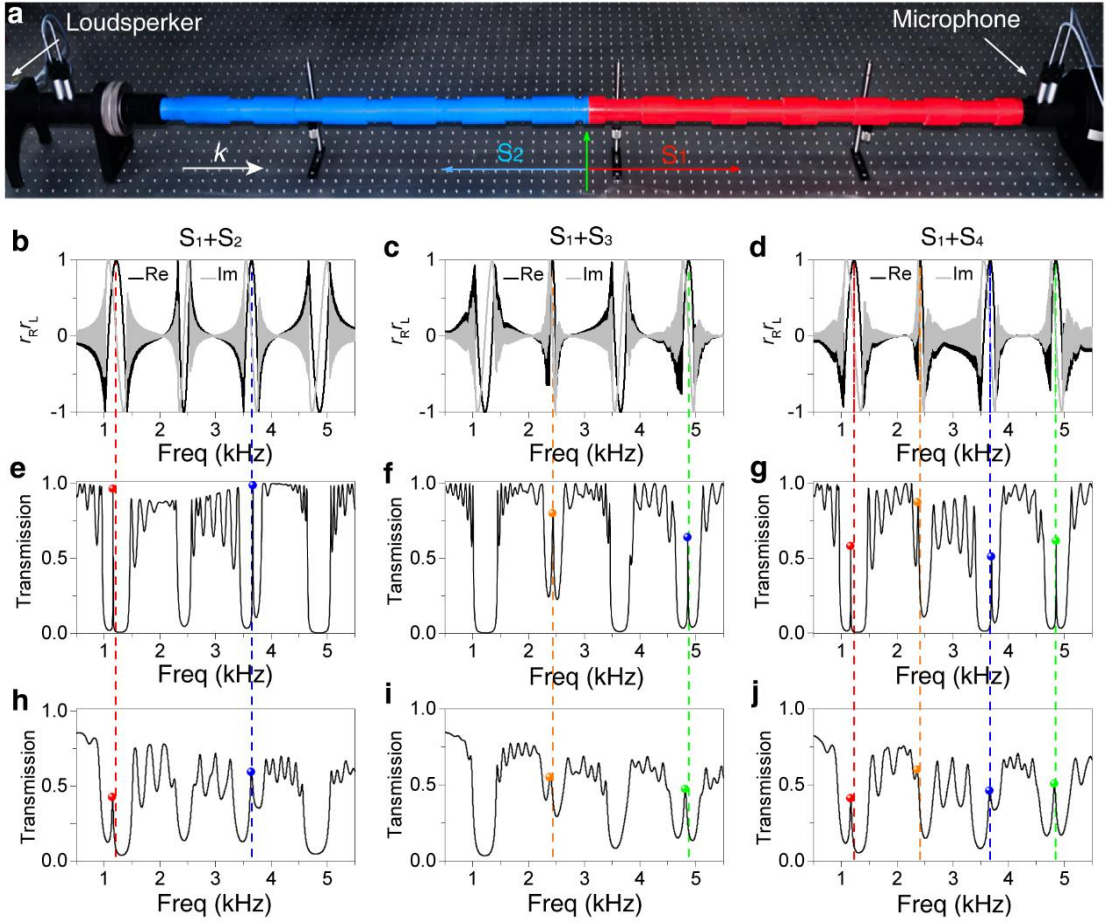

**Figure 5| Experimental set-up and observation of the designed interface states. a**, Photo of experimental set-up. Two acoustic PCs ($S_1$ in red and $S_i$ in blue) are connected at the junction marked by green arrow. **b-d**, The product of the reflection coefficients $r_{s1}r_{si}$ for the combinations of $S_1$ and $S_i$ (i=2,3,4), respectively. The frequencies where $r_{S1}r_{Si}=1$ are marked by colored dashed lines in each bandgap. **e-g**, The simulated transmission spectra of two connected PCs (6 unit cells for both the left and right PCs), for $S_1+S_2$, $S_1+S_3$, and $S_1+S_4$, respectively. **h-j**, The corresponding measured transmission spectra, where the transmission peaks arising from the interface states in each bandgap are highlighted by colored dots.

Further, the existences of topological interface states are experimentally verified, and the experimental set-up is shown in Fig. 5a. The sample contains two connected PCs, the right-hand one in red is manufactured with the geometric parameters of $S_1$ and the left-hand one in blue is manufactured with the geometric parameters of $S_{2,3,4}$ (only the photo of $S_1+S_2$ is shown). The junction is marked by the green arrow, and the PCs on either sides are truncated at the boundaries of an intact unit cell. Meanwhile, a loudspeaker and four microphones are used to generate input signals and to measure the transmission spectra, respectively. The measured transmission spectra for the three configurations are respectively shown in the lowest panel of Fig. 5. The transmission peaks marked by colored dots in the corresponding bandgaps confirm the existence of the interface states. The experimental results verify that interface states appear in all odd bandgaps for configurations of $S_1+S_2$, whereas in all even bandgaps for configurations of $S_1+S_3$, moreover, in all bandgaps for configurations of $S_1+S_4$. The frequencies of interface states agree well with our theoretical predictions and simulated results.

For the configuration of $S_1+S_4$, the simulated acoustic pressure distributions for the four interface states within each bandgap are shown in Figs. 6a-6d, corresponding to the frequencies of transmission peaks, 1.170 kHz, 2.381 kHz, 3.698 kHz, and 4.858 kHz, respectively. It can be observed that the pressure fields of these interface states are localized at the vicinity of the interface ($x$=0) and decay exponentially away from the interface. Meanwhile, we also measured the spatial distributions of the pressure fields at the frequencies of the interface states (see the details for experiments in Method and Supplementary Note 5). The measured pressure fields |$P$| at the four transmission peaks in four bandgaps are plotted in different colored dots in Figs. 6e-6h (complete measured data are available in Supplementary Figures 4b-4e). Excellent agreements between experiments and simulations (black curves) can be observed.

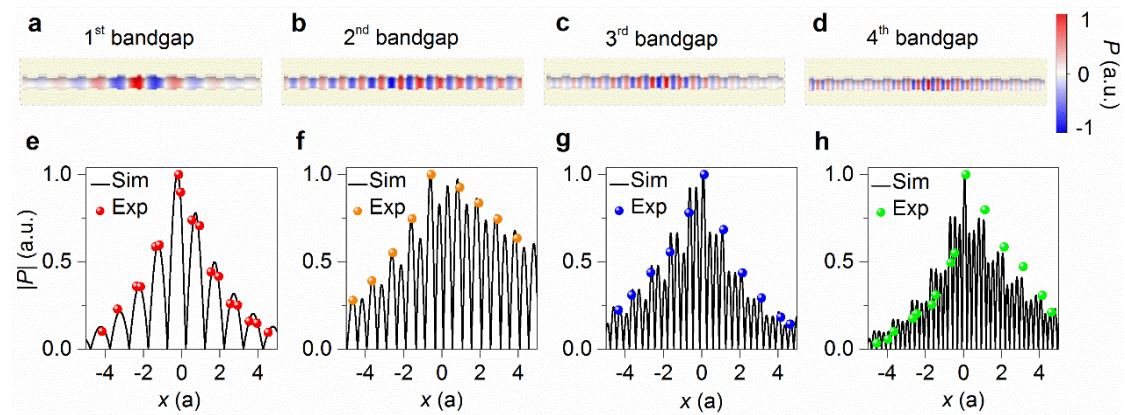

**Figure 6| Pressure fields of the interface states. a-d**, The simulated distributions of

acoustic pressure (real part of $P$) for the four interface states existing simultaneously in the $S_1+S_4$ configuration. **e-h**, Experimental measurements of the pressure field $|P|$ of the 4 interface states (colored dots). $|P|$ is normalized and the corresponding numerical results are plotted by the black curves.

**Discussion**

In summary, we have proposed a scheme to design interface states at the junction of two 1D PCs. The topological phases of bandgaps for each PC can be modulated by geometric parameters. The full phase diagrams of every bandgap in the parameter space can provide complete information on realizing interface states in arbitrary bandgap or bandgaps. As examples, we exhibited how to select the parameters of the connected PCs based on the phase diagrams for realizing interface states within a solo bandgap, within all odd or even bandgaps, as well as within all bandgaps. We also constructed real systems of connected PCs and measured transmission peaks in the bandgaps as well as the corresponding spatial field profiles. The results confirm the existences of interface states theoretically predicted from the phase diagrams. Moreover, this description of full phase diagrams can be extended to design topological interface states in other kinds of periodic systems, such as photonic crystals and designer plasmonic crystals.

**Acknowledgements**

We thank Prof. C. T. Chan, Dr. M. Xiao and A. Chen for helpful discussions. This work is supported by National Natural Science Foundation of China (NSFC) (11574037 and 91750102), and the Areas of Excellence Scheme grant (AOE/P-02/12) from Research Grant Council (RGC) of Hong Kong. D.Z.H. is also funded by the open projects of Key Laboratory of Micro- and Nano-Photonic Structures (Ministry of Education).

**Author Contributions**

Y. M., D. H., and W. W. conceived the original idea. Y. M., R. Z., X. W. and X. L. performed the analytical and numerical calculation. Y. H. and H. X. supported the fabrication process of the sample. Y. M., X. W. and X. L. carried out the experiments. Y. M., X. W., R. Z., P. H., L. G. and D. H. analyzed the data, prepared the figures and wrote the manuscript. W. W., S. W. and D. H. supervised the project. All authors contributed to scientific discussions of the manuscript.

**Methods**

**Numerical Calculations of Zak phase.** We use a discretized form of Eq. (3) to calculate the Zak phase:

$$\theta_n^{Zak} = -\text{Im} \sum_{j=1}^{N-1} \ln \left[ \frac{1}{2\rho v^2} \int_{\text{unit cell}} d\mathbf{r} dx u_{n,k_j}^*(x,\mathbf{r}) u_{n,k_{j+1}}(x,\mathbf{r}) \right], \quad (5)$$

where $u_{n,k}(x,\mathbf{r})=P_{n,k}(x,\mathbf{r})\exp(-ikx)$ is obtained numerically by normalizing the pressure eigenfunction with the orthogonal relationship $\int_{\text{unit cell}} d\mathbf{r} dx (1/2\rho v^2) |P_{n,k}(x,\mathbf{r})|^2 = 1$. Eq. (5) gives the Zak phase precisely in the limit $k_j=k_{j+1}-k_j \to 0$, where the summation of $k_j$ ranges from $-\pi/a$ to $\pi/a$. The periodic gauge, $P_{n,k=\pi/p}(x,\mathbf{r})=P_{n,k=-\pi/k}(x,\mathbf{r})$, is adopted in the calculation.

**Numerical calculations.** Numerical simulations are performed using the commercial finite element method software COMSOL Multiphysics. Eigenmode calculations are performed for calculating the band structures and finding eigen-field. Floquet periodic boundary conditions are imposed on periodic surfaces of the unit cell along the *x*-axis while sound hard boundary conditions are applied on all the other surfaces, the background surroundings and filler of the tubes is air (mass density $\rho$=1.3 kg·m$^{-3}$, the speed of sound $v$=343 m·s$^{-1}$). Frequency domain calculations are used to calculate the pressure response spectra.

**Sample fabrication.** The unit cells of acoustic waveguides are fabricated by a fused deposition modeling (FDM) 3D printer (Tiertime Technology Co., Ltd) with a kind of hard plastic Acrylonitrile Butadiene Styrene (ABS) plastics, and then are connected with super glue. The joints between unit cells are further sealed with vacuum grease to minimize loss. The wall of tubes is rigid and thick enough (~ 0.3 cm) to be regarded as sound hard boundaries.

**Experiment set-up and measurement.** The transmission spectra are measured in the standard method with two Brüel and Kjær type-4206 impedance tubes fix on either side of the sample. Plane waves are generated by a loud speaker and entered the waveguide from the far end of the left tube. The frequencies of the plane wave range from 0 to 6.4 kHz. Four microphones (Brüel and Kjær type-2670) are used to detect the incident, reflected and transmitted waves. Two of them are fixed at impedance tube on the left hand to detect the incident and reflected signals, and the other two are fixed at the impedance tube on the right hand to detect the transmitted signals.

For measuring the spatial pressure distribution, 80 small holes are opened on the sample of $S_1+S_4$, and a side hole near the interface of sample is opened as well (Supplementary Fig. 4a). The acoustic waves are generated by the loud speaker, and entered the waveguide from the side hole, the source can be treated as point source in our experiments (Supplementary Note 5). When measure the pressure distribution of the interface states, both ends of the sample are bunged up with sound-absorbing sponge. Two microphones are fixed at impedance tube to detect the incident and reflected signals, and the other two microphones move from the leftmost end to the rightmost end to detect the pressure of each hole. The probe holes are sealed with rubber plugs when they are not under measurement.

**Data availability.** The data which support the figures and other findings within this paper are available from the corresponding authors upon request.

## Additional Information

Supplementary information is available in the online version of the paper.

## Competing Financial Interests

The authors declare no competing financial interests.